\documentclass[12pt,letterpaper,onecolumn]{article}
\usepackage[latin1]{inputenc}
\usepackage{amsmath}
\usepackage{amsfonts}
\usepackage{amssymb}
\usepackage{braket}
\usepackage[left=1.00in, right=1.00in, top=1.00in, bottom=1.00in]{geometry}
\author{Horace P. Yuen\\
Department of Electrical Engineering and Computer Science\\
Department of Physics and Astronomy\\
Northwestern University, Evanston Il. 60208\\
yuen@eecs.northwestern.edu
}
\title{Unconditional Security In Quantum Key Distribution}
\begin{document}
\maketitle
\begin{abstract}
\textit{It has been widely claimed and believed that many protocols in quantum key distribution, especially the single-photon BB84 protocol, have been proved unconditionally secure at least in principle, for both asymptotic and finite protocols with realistic bit lengths. In this paper it is pointed out that the only known quantitative justification for such claims is based on incorrect assertions. The precise security requirements are described in terms of the attacker's sequence and bit error probabilities in estimating the key. The extent to which such requirements can be met from a proper trace distance criterion is established. The results show that the quantitative security levels obtainable in concrete protocols with ideal devices do not rule out drastic breach of security unless privacy amplification is more properly applied, while it is problematic whether a positive net key can be generated from current approaches.}
\end{abstract}

In quantum key distribution (QKD), quantum effects that have no classical analog are utilized for generating a sequence of bits (the secrey key $K$) between two parties A and B that are known only to themselves. The typical approach involves information-disturbance tradeoff in BB84 type protocols [1], but other quantum approaches without using such a tradeoff is possible, say in KCQ (keyed communication in quantum noise) [2]. It has been claimed since long ago and maintained to this day [3] that BB84 has been proved to possess \textit{unconditional security} (UCS), which is in fact the major advantage of QKD compared to other known ciphers. What does UCS mean exactly?

In conventional classical key distribution such as the public key RSA scheme to which QKD is often compared to, security is based on computational complexity that it is computationally difficult for an adversary E to compute the key though it is in principle possible. This means $K$ does not possess information-theoretic security (ITS), that there is no intrinsic probabilistic uncertainty to $K$. In this paper we assume the cryptosystem model is a complete representation of all the relevant physical attributes of the cryptographic situation, although the fact that it has not been in BB84 is a major loophole of concrete protocols [4]. Thus, UCS is to be discussed under the assumption that everything fits the ideal system model, as such a security claim is usually so understood in the literature.

It is evident from the above description that UCS means no more than ITS for all possible uses of $K$, assuming the laws of quantum mechanics are universal. This in turn means UCS is a \textit{quantitative} issue since it involves probability (as in fact quantum mechanics itself does) and so the numerical value of the probability of E's success in finding $K$ gives one actual (unconditional) security level of the QKD protocol. Indeed, E may want to identify only a portion of $K$, so her probabilities of finding various subsets $K^*$ of $K$ are also important quantitative criteria associated with UCS. In addition, when $K$ is used in one-time pad form (xor into the data bits), as often suggested for QKD to get true UCS instead of using $K$ as the seedkey of a conventional cipher such as AES, the number of actual bit errors E makes in estimating $K$ or its subsets from her attacks, to be called E's bit error rates (BER) in contrast to the above ``sequence error probabilities'', would be relevant additional quantitative criteria for UCS. Such leak would be equivalent to a leak from a nonuniform a priori probability distribution on $K$.

One more major distinction needs to be made, raw security versus KPA security [5]. E can try to estimate the above probabilities from just the probe she set and the public exchange before $K$ is actually used. The quantitative results she so obtained give the \textit{raw security} of $K$. When $K$ is actually used, E may obtain further information and she could in principle make measurements on her probe after such information becomes available. The resulting probabilities determine the ``composition security'' of $K$. We restrict to a specific form of composition security that E could readily launch in many applications, known-plaintext attacks (KPA). Indeed we would restrict to just KPA where a segment of $K$ is known to E exactly, say from knowing some data bits and of course the ciphertext bits when $K$ is used in one-time pad form. Such partial knowledge of $K$ may help her determine the rest of $K$ and hence the rest of the data segment she did not know. \textit{KPA security} refers to these quantitative probabilities E may get. Note that ITS in raw security is obtained in conventional key expansion [5] from a shared secret key which is also needed in QKD for message authentication, but there is no IPS under known plaintext attacks.

The ideal UCS or ITS is obtained when E has a uniform probability $U(k)$ for all the possible values of the $n$-bit generated key $K$ and it is independent with whatever information E may possess. It would be good UCS if such a situation is obtained with a sufficiently high probability. This is precisely the claim that has been maintained in the QKD literature since [6,7] to the recent review in [3] and beyond. We will also show the other mathematically unspecific justifications of UCS in terms of ``distinguishability advantage'' is not applicable. Both of these justifications are given by a trace-distance criterion $d$. In this paper we will determine the extent $d$ could provide quantitative UCS.

Before proceeding, it may be \textit{noted} that this issue lies at the heart of the whole security foundation of QKD, of exactly what security at what level with what empirical meaning one can obtain from QKD. In contrast to most issues in physics, this cannot be decided by an experiment and a careful conceptual and mathematical development is the only way to resolve it.

Let $K^{*}$ be a subset of $K$ from an arbitrary fixed subset of the $n$ bit positions of $K$. Thus $K^{*}$ contains $1$ to $n$ bits and may take one of $2^{|K^{*}|}$ possible values. Let $p_{1}(K^{*})$ be E's optimal probability of estimating $K^{*}$  from her attack. The probability $p_{1}(K)$ is especially important because it is the probability of E successfully estimating the whole $K$. For raw security one needs to upper bound each $p_{1}(k^{*})$ to an acceptable level, say \begin{equation} \label{eq1}
p_{1}(k^{*}) \leq 2^{-|K^{*}|}+\epsilon'
\end{equation} for some $\epsilon'$ that may depend on $|K^{*}|$. This may happen only with a certain probability itself depending on the exact value of $K^{*}$ and other system parameters. It is usually only possible to usefully bound the average $\overline{p_{1}}(K^{*})$ over the values of $K^{*}$, which replaces the individual value in the left side of \eqref{eq1}. Such a bound can be converted to the form of \eqref{eq1} by application of Markov Inequality.

Under KPA, E knows a subset $X_{1}$ of  the data $X$ encrypted by $K$. In the one-time pad format E would then know a corresponding segment $K_{1}=k_{1}$ of $K$ which she could use to help her get other subsets $K_{2}^*$ of $K_{2}$ in the rest of $K=k_1\bigcup K_2$. For UCS one needs to bound, for small $\epsilon''$ that may depend on $|K_{1}|$ and $|K_{2}|$, \begin{equation} \label{eq2}
p_{1}(k_{2}^{*}|K_{1}=k_{1}) \leq 2^{-|K_{2}^{*}|}+\epsilon''
\end{equation} when a portion $K_{1}$ of $K$ is known to be $k_{1}$ and a subset $K_{2}^{*}$ is to be estimated. Again, an average over $K_{1}$ and $K_{2}^*$ may be needed to derive such bounds. \textit{Note} that for a uniform key, \eqref{eq1} would be satisfied with equality for $\epsilon'=0$, and if it is independent of E's information, \eqref{eq2} would be similarly satisfied with $\epsilon'' = 0$, thus giving perfect UCS. If such a situation can be obtained with high probability (from other random parameters in the system), then the protocol has perfect UCS with a high probability, which is exactly the current claim [3,6,7].

Note that the criteria of \eqref{eq1}-\eqref{eq2} are the only operational meaningful security criteria that any other criterion in the form of an information theoretic quantity [8] must reduce to, including mutual information and variational distance. This should be clear if one asks the question: so what is the empirical or operational guarantee given the criterion is at a given level?

The claim that $K$ gives the above perfect UCS with a high probability is made on behalf of the trace distance criterion $d$ defined as follows. During key generation E sets her probe and the protocol goes forward.  After privacy amplification the final key $K$ is generated with corresponding ``prior probability'' $p(k)$ and probe state $\rho^k_E$ on each $k$. Let \begin{equation} \label{eq3}
\rho = \sum_k p(k)\ket{k}\bra{k}
\end{equation} for $N$ orthonormal $\ket{k}$'s in space $\mathcal H_K$, $N=2^n$. Let $\rho_E = \sum_k p(k) \rho_E^k$, $\rho_{KE} = \sum_k p(k)\ket{k}\bra{k}\otimes \rho_E^k$. The criterion $d$ is defined to be \begin{equation} \label{eq4}
d \equiv \frac{1}{2}\parallel\rho_{KE}-\rho_U\otimes\rho_E\parallel_1
\end{equation} where $\rho_U$ is given by \eqref{eq3} with $p(k)=U(k)$ for the uniform random variable $U$. It can be readily shown (similar to Lemma 2 in [6]) that \begin{equation} \label{eq5}
d = \frac{1}{2}\sum_k \parallel p(k)\rho_E^k-\frac{1}{N}\rho_E\parallel_1
\end{equation} A key $K$ with $d\leq\epsilon$ is called ``$\epsilon$-secure'', as it has been forced by privacy amplification to be $\epsilon$-close to $U$. But what is the operational meaning of $d\leq \epsilon$?

The major interpretation that has been given to $d\leq \epsilon$ amounts to saying perfect UCS is obtained with a probability $\geq 1-\epsilon$. In [6] it is explicitly stated ``The real and the ideal setting can be considered identical with probability at least $1-\epsilon$.'' In [9,3] it is expressed with a different nuance with $\epsilon$ understood as ``maximum failure probability'' of the protocol  ``where 'failure' means that 'something went wrong', e.g., that an adversary might have gained some information on K''.

The justification of such erroneous interpretation of $d$ is derived from the interpretation of Lemma 1 in [6] that the variational distance $v(P,Q)$ between two distributions $P$ and $Q$ on the same sample space, the classical counterpart of trace distance, ``can be interpreted as the probability that two random experiments described by $P$ and $Q$, respectively, are different.'' That this interpretation cannot be true in any situation has been discussed in [5,10]. Here we give a simple example to bring out why.

Consider the distribution upon a measurement result with $P_i=\frac{1+2\epsilon}{N}$ for $i\in \overline{1-\frac{N}{2}}$ and $P_i=\frac{1-2\epsilon}{N}$ for $i\in \overline{(\frac{N}{2}+1)-N}$, so that $v(P,U)=\epsilon$. Then E ``gains information'' compared to the ideal case with probability 1/2, not $\epsilon$. This example clearly shows that variational distance is not the maximum probability that information is leaked.

Operational security significance for $d$ can be derived, however, from the classical properties of the variational distance between $K$ and the uniform distribution $U$. Under $d\leq \epsilon$, condition \eqref{eq1} holds for the $K^*$-averaged $p_1(K^*)$ with $\epsilon'=\epsilon$. We separate out the case for the whole $K$ due to its crucial role \begin{equation} \label{eq6}
\overline{p}_1(K) \leq \frac{1}{N} + \epsilon
\end{equation} Under KPA, it is not possible to have $p_1(K^*_2|K_1=k_1)$ lower bounded by a small number because it can be arbitrarily close to 1 for any given $K_1=k_1$. Such $k_2$ can occur with arbitrarily small but nonzero $p(k)$ to satisfy any $d\leq \epsilon$ constraint for nonzero $\epsilon$. The best one can hope for is a bound \eqref{eq2} when $K_1$ itself is averaged over. This in fact holds under $d \leq \epsilon$ where the $K_{1}$ and $K_{2}^*$ averaged $\overline{p}_{1}$ satisfies \begin{equation}
\overline{p}_1(K_2^*|K_1) \leq 2^{-|K_2^*|} + \epsilon
\end{equation} We outline here the proof of (7) which covers the raw security of no conditioning as a special case. Let $Y$ be the measurement random variable of the relevant optimum quantum measurement E makes. One can write,
\begin{equation*} \label{eq7} \overline{p}_1(K_2^*|K_1)=\displaystyle\sum\limits_{k_1} p(k_1) \cdot \displaystyle\sum\limits_{k_2^*} p(k_2^*|k_1)p(k_2^*|k_1) \cdot \displaystyle\sum\limits_{y\epsilon I_{k_2^*|k_1}} p(y)p(k_2^*|yk_1)
\end{equation*} where $I_{k_2^*|k_1}$ is the optimal decision region on $K_2^*$ given $K_1=k_1$, irrespective of $K_2'$. From equation (11.137) of [11] it follows that $p(k_2^*|yk_1)\leq p(k_2^*|k_1)+\epsilon_y$ with $\displaystyle\sum\limits_y p(y)\epsilon_y = \epsilon$. Extending the sum in $y \epsilon I_{k_2^*|k_1}$ over all $y$ leads to (7).

Inequality (6) was previously given in [10], the full operational significance of $d \leq \epsilon$ is given here in (7) for the first time. These sequence error probabilities constitute the appropriate criteria when $K$ is used as the seed key in a conventional cipher such as AES. For the more commonly suggested use of $K$ in one-time pad form, the bit error rate (BER) is also important because E may get many bits correctly even when she gets the whole $K^*$ wrong. This is the common distinction between sequence error rate and bit error rate in ordinary communications. BER is defined to be the per bit error probability, with $N^*=2^{|K^*|}$, \begin{equation} \label{eq10}
p_b \equiv P_b(K) = \frac{1}{N^*}\sum^{N^*}_{i=1}P_e(i)
\end{equation} where $P_e(i)$ is the probability that the $i$th bit in $K^*$ is incorrectly obtained from Eve's estimate of $K^*$. Here we summarize the BER result in [11].

The only known general lower bound on $P_b$ is the Fano Inequality [11], which gives in this case, with $I_{ac}$ being E's quantum accessible information, \begin{equation} \label{eq11}
n \mathcal H(p_b) \ge H(K)- I_{ac}
\end{equation} where $\mathcal H(\cdot)$ is the binary entropy function and $H(K^*)$ the entropy of $K^*$. The $H(K)$ for $K$ is determined by $p(k)$ in (3). From $d$ we can bound $H(K)$ by [11, theorem 17.3.3] which yields, for $p_b = \frac{1}{2}-\epsilon^\prime$  and small $\epsilon^\prime$, $\epsilon^\prime \leq (\epsilon/{4\log e})^{1/2}$. Since Markov Inequality needs to be used twice before this $\epsilon'$ is applied, it is similar to the case of using it three times and the final $\epsilon'$ in \eqref{eq1} is thus \begin{equation} \label{eq12}
\epsilon^\prime   \leq \epsilon^{\frac{1}{4}}/2\sqrt{\log e} 
\end{equation} As expected, the BER guarantee from (11) is worse than that of the corresponding sequence error probability.

In contrast to the sequence error case, there is no result similar to(6)-(7) for subsets $K^*$ or $K_2^*$ because there is no lower bound on $H(K^*)$ or $H(K_2^*|K_1)$ from $d$ or $H(K)$, and it is possible to have arbitrarily small but nonzero $H(K^*)$ especially when conditioned on $K_1=k_1$. Thus, the result on BER is limited to E's attack on the whole $K$ in raw security.

There is an original argument [15] that purports to show $d$ has general raw and composition security from its mere form of \eqref{eq4}, because the optimum binary quantum decision probability $P_c$ between two states $\rho_o$ and $\rho_1$ with a priori probability $P_o$ and $P_1$ is given by \begin{equation} \label{eq8}
P_c = \frac{1}{2} + ||P_o\rho_o - P_1\rho_1||_1
\end{equation} and the two terms in \eqref{eq4} represent the real and the ideal situation, thus $d$ provides a bound on the ``distinguishability advantage''. However, E is not trying to distinguish the two situations by a binary decision, thus (4) and (12) give the wrong criterion in either raw or KPA security. The correct criteria are (1)-(2) in terms of E's probability of success in identifying various $K^*$. Note also that \eqref{eq4} is itself a fictitious representation and in any case not available to E, or she could just measure on $\mathcal H_K$ to get $K$. The form \eqref{eq5} for $d$ is much less misleading than the entanglement form \eqref{eq4}. Further discussion can be found in [10].

That ``universal composable security'' does not follow from (4) and (12) is especially clear in the case of BER where no bound on $p_b(K_2^*|K_1)$ can apparently be derived from $d \leq \epsilon$, due to the very nonlinear relation between $p_b$ and $d$ already apparent in (10). To establish a security claim, one needs to write down mathematically what is being claimed and provide a derivation from given, in this case $d \leq \epsilon$. The incorrect ``maximum failure probability'' interpretation of $d$ gives such a derivation for raw and composition security, but it cannot be true. We have provided the correct security guarantee (6)-(7) and (10)-(11) from $d \leq \epsilon$, but they are far weaker than those from the incrorect interpretations [12].

The significant point in this correction is that E makes an $N$-ary decision in estimating $K$, or an $N^*$-ary decision in estimating $K^*$.  From the viewpoint of a binary decision for (4) and (12) with $P_0 = \frac{1}{2}$, $d=2^{-10}$ may appear sufficient. However, for an $N$-ary decision with, say $n=1,000$, it follows from \eqref{eq6} that such a $d$ value does \textit{not} rule out a \textit{disastrous breach} of security: that the whole 1,000 bits key may be obtained with a 0.1\% probability. It is clear the problem is one needs to look at the \textit{quantitative} security level with respect to a proper \textit{reference level}.

There is the persistent intuition that a criterion should be fine if the level is brought down to a sufficiently small value, assuming the value is zero in the ideal case. This is true if the value is exactly zero, but the whole question is how small is sufficiently small, or what the reference level is. It is a \textit{quantitative} issue through and through, UCS does not imply security if its level is not good. In this connection, it may be pointed out that  $I_{ac}$ has been used as the QKD security criterion from the beginning till some work to date. It has been largely abandoned in the theory literature because it does not rule out possible disastrous leaks from quantum information locking against KPA when E has quantum memory [13, 14]. Indeed, the incorrect interpretation of $d$ was proposed [13] in place of [15] for exactly such problem. On the other hand, it can be shown [16] that if $I_{ac}$ is small enough such locking information cannot be utilized either. A good reference level for this with an $n$-bit key is $d=2^{-n}$.

The raw security guarantee (6) from $d$ [17] is totally inadequate for the analyzed finite protocols with their numerical values of the parameters, as follows. The most up to date finite-key analysis of the single photon BB84 protocol with no loss and ideal devices [18] gives typical $d$ levels of $10^{-9}$ for 5\% QBER and 10\% key rate, for $n\sim 10^5$ at the limit of present day error correcting code block length. After MAI is applied twice for $K$ and privacy amplification averaging, the resulting individual probability guarantee with $d^{\frac{1}{3}}$ is $10^{-3}$. That is, it is \textit{not} ruled out that Eve may have an estimate that has a probability of 0.001 of finding the whole 10,000 bit key, a \textit{disastrous} breach of security. In such case, there is effectively only a 10 bit protection of the 10,000 bit key. The BER guarantee of (9)-(11) shows E has $p_b\sim 0.49$ instead of 0.5 when attacking the whole $K$, which for $n=10^5$ amounts to knowing 1,000 bits more, considerably bigger than a favorable (to E) binomial fluctuation level of $\sim 200$. In the NEC experimental decoy state system [19] the criterion $I_{ac}$ was used but a corresponding $\overline{p}_1(K)$ is also given [20] consistent with the result of [2], with $\overline{p}_1(K) \sim 10^{-6}$ for $n \sim 4,000$. After a cube root to $d$ this implies the probability guarantee is the way too large 1\% with BER $p_b \sim 0.4$ error probability. These results show that a much smaller $d$ value needs to be guaranteed in privacy amplification. See [21].

In conclusion, we have specified the operational requirement of unconditional security in cryptography and determined the extent it can be satisfied by the trace distance criterion $d \leq \epsilon$. It is seen that the $d$ values given in the literature for finite protocols are very far from ruling out possible drastic breach of security. In addition, the results point to a serious gap in the security proofs in connection with current treatments of error correction, and as a consequence no concrete full protocol has been proved secure even under just collective attacks [21]. It appears radically new elements need to be introduced to make QKD provably secure with meaningful levels of security and key rate.

\section*{Acknowledgements}
This work was supported by the Air Force Office of Scientific Research.


\begin{thebibliography}{9}
\bibitem{ref12} Note also that under the incorrect interpretation of $d$, Markov Inequality needs to be applied only once from average over privacy amplification codes to arrive at (1)-(2) for $\epsilon = 0$ with mobility $\geq 1-d$.
\bibitem{ref17}Condition (6) can be satisfied with equality [2].
\end{thebibliography}
\end{document}